\documentclass[onecolumn,superscriptaddress,secnumarabic,amssymb,amsmath,nobibnotes,aps,prd,nofootinbib,altaffilletter]{revtex4}
\usepackage{amsmath,amssymb}
\usepackage{graphicx}
\usepackage{dcolumn}
\usepackage{bm,color}
\usepackage{hyperref}
\usepackage{accents}
\usepackage{amssymb,float}
\usepackage{amsmath}
\usepackage{multirow}
\usepackage{siunitx}
\usepackage{tabularx}
\usepackage{booktabs}
\usepackage{url}

\DeclareSIUnit\parsec{pc}

\newcommand{\udt}[3]{#1^{#2}_{\phantom{#2}#3}}
\newcommand{\udut}[4]{#1^{#2\phantom{#3}#4}_{\phantom{#2}#3\phantom{#4}}}

\newcommand{\dut}[3]{#1_{#2}^{\phantom{#2}#3}}

\newcommand{\lc}[1]{\accentset{\circ}{#1}}%Levi-Civita connection

\begin{document}

\title[\texorpdfstring{$f(T,B)$}{} Gravity in the late Universe in the context of local measurements]{\texorpdfstring{$f(T,B)$}{} Gravity in the late Universe in the context of local measurements}

\author{Rebecca Briffa}
\email{rebecca.briffa.16@um.edu.mt}
\affiliation{Institute of Space Sciences and Astronomy, University of Malta, Msida, Malta}
\affiliation{Department of Physics, University of Malta, Msida, Malta}

\author{Celia Escamilla-Rivera}
\email{celia.escamilla@nucleares.unam.mx}
\affiliation{Instituto de Ciencias Nucleares, Universidad Nacional Aut\'{o}noma de M\'{e}xico, Circuito Exterior C.U., A.P. 70-543, M\'exico D.F. 04510, M\'{e}xico}

\author{Jackson Levi Said}
\email{jackson.said@um.edu.mt}
\affiliation{Institute of Space Sciences and Astronomy, University of Malta, Msida, Malta}
\affiliation{Department of Physics, University of Malta, Msida, Malta}

\author{Jurgen Mifsud}
\email{jurgen.mifsud@um.edu.mt}
\affiliation{Institute of Space Sciences and Astronomy, University of Malta, Msida, Malta}
\affiliation{Department of Physics, University of Malta, Msida, Malta}

%%%%%%%%%%%%%%%%%%%%%%%%%%%%%%%%%%%%%%%%%%%%%%%%
%%%%%%%%%%%%%%%%%%%%%%%%%%%%%%%%%%%%%%%%%%%%%%%%

\begin{abstract}
We explore the viability of three models in $f(T,B)$ gravity using data from recent surveys based on cosmic chronometers, the Pantheon data set, and baryonic acoustic oscillation data. We also assess the consistency of these models and data set combinations with two important priors on the Hubble constant coming from the SH0ES Team and measurements using the tip of the red giant branch respectively. These give the highest and lowest values of the Hubble constant coming from cosmology independent studies. In general, our analysis does provide a more consistent fit for the late time data being analyzed. However, each model does include an additional model parameter in comparison with the concordance model. We close the analysis with a comparative analysis in which each model, data set and Hubble constant prior combination are cross-analyzed against each other.
\end{abstract}

%%%%%%%%%%%%%%%%%%%%%%%%%%%%%%%%%%%%%%%%%%%%%%%%
%%%%%%%%%%%%%%%%%%%%%%%%%%%%%%%%%%%%%%%%%%%%%%%%

\maketitle

%%%%%%%%%%%%%%%%%%%%%%%%%%%%%%%%%%%%%%%%%%%%%%%%
%%%%%%%%%%%%%%%%%%%%%%%%%%%%%%%%%%%%%%%%%%%%%%%%

\section{\label{sec:intro}Introduction}

The recent indication of a statistical tension \cite{Bernal:2016gxb,DiValentino:2020zio,DiValentino:2021izs,Abdalla:2022yfr, Dainotti:2021pqg, Dainotti:2022bzg} in value of the Hubble constant $H_0$ has prompted a reassessment of possible modifications beyond standard vanilla $\Lambda$CDM. While the $H_0$ problem may be the result of systematics, the mismatch in reported values of the Hubble constant has been reported by a number of surveys indicating a more serious nature to the issue. On the other hand, this standard model of cosmology can produce a consistent evolution of the Universe \cite{Planck:2018vyg}. However, this is only possible with the inclusion of exotic particles physics using cold dark matter (CDM) \cite{Baudis:2016qwx,Bertone:2004pz} which govern the dynamics of galaxies, and a cosmological constant \cite{Peebles:2002gy,Copeland:2006wr}, which is the source of the late time accelerating Universe \cite{SupernovaSearchTeam:1998fmf,SupernovaCosmologyProject:1998vns}. Despite great efforts to directly detect CDM \cite{Gaitskell:2004gd} and to formulate a deeper explanation for the cosmological constant \cite{Weinberg:1988cp}, these prospects remain elusive.

The $H_0$ tension problem has led to increased efforts to determine the current value of the Hubble parameter in order to better understand the degree of disparity between measurements that are independent of cosmological models and those that require a cosmological model to make a Hubble constant estimate. The most prominent example of the former is the latest announced result of the SH0ES Team \cite{Scolnic:2021amr,Brout:2021mpj,Riess:2021jrx} which is based on observations of supernovae type Ia (SNe Ia) with calibration coming from Cepheid measurements. This gives a Hubble constant of $73.3 \pm 1.04 \,{\rm km\, s}^{-1} {\rm Mpc}^{-1}$ which is the result of data points taken below $z=1$. A relatively high value of the Hubble constant is also supported by the H0LiCOW Collaboration \cite{Wong:2019kwg} which is based on six strong lensing measurements and gives a value of $73.3^{+1.7}_{-1.8} \,{\rm km\, s}^{-1} {\rm Mpc}^{-1}$. 
Other model independent measurements do point to lower values of the Hubble constant such as measurements of the tip of the red giant branch who give a value of $ 69.8\pm 1.7 \,{\rm km\, s}^{-1} {\rm Mpc}^{-1}$ \cite{Freedman:2019jwv,Freedman:2021ahq}, 
and measurements from novel approaches of gravitational wave standard sirens \cite{LIGOScientific:2017adf} which gives a higher uncertainty result of $70.0^{+12}_{-8} \,{\rm km\, s}^{-1} {\rm Mpc}^{-1}$. On the other hand, model dependent measurements based on early Universe measurements give a drastically lower value of the Hubble constant. The most prominent of these measurements is the Planck Collaboration result of $67.4 \pm 0.5 \,{\rm km\, s}^{-1} {\rm Mpc}^{-1}$ and the Atacama Cosmology Telescope (ACT) Hubble constant $67.9\pm 1.5 \,{\rm km\, s}^{-1} {\rm Mpc}^{-1}$ \cite{ACT:2020gnv}.

One direction that has received heightened community effort in the literature is that of modifications to general relativity (GR) where numerous proposals have been made to better meet the observational constraints of recent surveys \cite{Sotiriou:2008rp,Clifton:2011jh,CANTATA:2021ktz}. By and large, most of these proposed gravitational models build on GR and extend the Einstein-Hilbert action to meet specific theoretical or phenomenological goals \cite{Faraoni:2008mf,Capozziello:2011et}. However, these expressions of gravity are mostly based on the same curvature-based geometric description of gravity as GR, which is sourced by the Levi-Civita connection \cite{misner1973gravitation,nakahara2003geometry}. Interestingly, other descriptions of geometric gravity are possible \cite{BeltranJimenez:2019esp}. One class of theories that has garnered growing interest in the literature is that of teleparallel gravity (TG) \cite{Bahamonde:2021gfp,Aldrovandi:2013wha,Cai:2015emx,Krssak:2018ywd}. This is based on the exchange of the curvature associated with the Levi-Civita connection with its teleparallel connection analogue which is curvature-less and gives an expression for torsion in gravity. Naturally, this implies that all measures of curvature will vanish so we must consider new tensors to define gravitational theories. One critical scalar that is central to TG is the torsion scalar $T$ which turns out to be equivalent to the regular curvature scalar $\lc{R}$ (over-circles represent quantities calculated with the Levi-Civita connection), up to a total divergence term. This means that we can define a \textit{Teleparallel equivalent of General Relativity} (TEGR) where the Einstein-Hilbert action is replaced with one based solely on the torsion scalar. The total divergence difference between the curvature and torsion scalars is denoted by $B$ and will be important for the work that follows. Indeed, it embodies the fourth order aspects of the curvature scalar which produce fourth order terms in many modifications of GR \cite{Lovelock:1971yv,Gonzalez:2015sha,Bahamonde:2019shr}.

In the TEGR context, the curvature scalar of the Einstein-Hilbert action is divided into the active torsion scalar $T$ and the total divergence contribution from $B$. Taking the same approach of many of the extended theories of gravity in curvature-based geometries, such as $f(\lc{R})$ gravity \cite{Sotiriou:2008rp,Faraoni:2008mf,Capozziello:2011et}, we can write an $f(T,B)$ action for the gravitational sector \cite{Bahamonde:2015zma,Bahamonde:2016grb,Paliathanasis:2017flf,Farrugia:2018gyz,Bahamonde:2016cul,Wright:2016ayu,Farrugia:2020fcu,Capozziello:2019msc,Farrugia:2018gyz,Escamilla-Rivera:2019ulu}. Models in $f(T,B)$ gravity have been explored at all levels of phenomenology, it hosts many models that are consistent with solar system level observations \cite{Farrugia:2020fcu}. In terms of gravitational waves, it predicts generically that they propagate at the speed of light but it does predict extra polarizations in comparison with GR \cite{Capozziello:2019msc,Farrugia:2018gyz}. Another important development was the introduction of exact and perturbed black hole solutions in Ref.~\cite{Bahamonde:2021srr}. In terms of cosmology, there have been a number of studies that probe both the background cosmological evolution of models in the theory \cite{Bahamonde:2016grb,Paliathanasis:2017flf,Bahamonde:2016cul,Escamilla-Rivera:2019ulu,Escamilla-Rivera:2021xql,Franco:2020lxx,Caruana:2020szx}, as well as its perturbative structure \cite{Bahamonde:2020lsm}. There has also been significant foundational work on the structure of the theory \cite{Wright:2016ayu} such as its exact relationship to $f(\lc{R})$ gravity \cite{Bahamonde:2015zma}.

The surge in surveys reporting values of the Hubble constant is the main driver of current efforts in modified gravity to define favoured models that may meet the observational challenges of the coming decade. In this work, we explore three models in $f(T,B)$ cosmology which have garnered popularity in the community. We also consider the impact that prior values on the Hubble constant may have on the resulting cosmological model parameter values. To do this, we first give an introduction to TG and its $f(T,B)$ formalism in Sec.~\ref{sec:intro_f_T}. We also describe its cosmology here, and then move onto information about the observational data we use and the $H_0$ priors in Ref.~\ref{sec:obs_data}. We give our main results in Sec.~\ref{sec:models} where we introduce the three models together with the statistical outputs of our analysis. Finally, we summarize our results in Sec.~\ref{sec:conc}.

%%%%%%%%%%%%%%%%%%%%%%%%%%%%%%%%%%%%%%%%%%%%%%%%
%%%%%%%%%%%%%%%%%%%%%%%%%%%%%%%%%%%%%%%%%%%%%%%%

\section{\label{sec:intro_f_T}\texorpdfstring{$f(T,B)$}{} FLRW Cosmology}

TG is based on the exchange of the curvature associated with the Levi-Civita connection \cite{Clifton:2011jh} with the torsion sourced through the teleparallel connection where torsion replaces curvature as the mediator of gravitation \cite{Bahamonde:2021gfp}. Curvature is a produce not of the metric tensor but of the Levi-Civita connection $\udt{\lc{\Gamma}}{\sigma}{\mu\nu}$ (over-circles are used throughout to denote quantities determined using the Levi-Civita connection), and so identical metric components can continue to result from a TG description of gravity. However, TG is different in that geometric torsion emerges from the teleparallel connection $\udt{\Gamma}{\sigma}{\mu\nu}$ which continues to satisfy metricity \cite{Hayashi:1979qx,Aldrovandi:2013wha}. As a note on the characteristic measures in gravitation, the regular Riemann tensor obviously does not vanish ($\udt{\lc{R}}{\beta}{\mu\nu\alpha} \neq 0$), but if the Levi-Civita connections are replaced with their teleparallel connection analogues then indeed curvature vanishes identically ($\udt{R}{\beta}{\mu\nu\alpha} = 0$) (see reviews in Refs. \cite{Krssak:2018ywd,Cai:2015emx,Aldrovandi:2013wha}), which points to the need for a different description of gravitational interactions.

Torsion is more conveniently expressed in terms of the tetrad $\udt{e}{A}{\mu}$ (and its inverses $\dut{E}{A}{\mu}$) whereby the metric becomes a derived quantity through the expressions
\begin{align}\label{metric_tetrad_rel}
    g_{\mu\nu}=\udt{e}{A}{\mu}\udt{e}{B}{\nu}\eta_{AB}\,,& &\eta_{AB} = \dut{E}{A}{\mu}\dut{E}{B}{\nu}g_{\mu\nu}\,,
\end{align}
where Latin indices represent coordinates on the tangent space while Greek indices continue to represent indices on the general manifold \cite{Cai:2015emx}. Tetrads are largely suppressed in GR but they do appear in some circumstances \cite{Chandrasekhar:1984siy}. Naturally, these tetrads must satisfy orthogonality conditions
\begin{align}
    \udt{e}{A}{\mu}\dut{E}{B}{\mu}=\delta^A_B\,,&  &\udt{e}{A}{\mu}\dut{E}{A}{\nu}=\delta^{\nu}_{\mu}\,,
\end{align}
for consistency.

Given the tetrad description, the teleparallel connection can directly be written as \cite{Weitzenbock1923,Krssak:2015oua}
\begin{equation}
    \udt{\Gamma}{\sigma}{\nu\mu} := \dut{E}{A}{\sigma}\left(\partial_{\mu}\udt{e}{A}{\nu} + \udt{\omega}{A}{B\mu}\udt{e}{B}{\nu}\right)\,,
\end{equation}
where the freedom of the theory is spread over the tetrad and the spin connection $\udt{\omega}{A}{B\mu}$ which is a flat connection, and thus represents the local Lorentz freedom, which is a result of the explicit appearance of the local indices in this description. Thus, the tetrad-spin connection pair embody the gravitational and Lorentz degrees of freedom of the theory which then appear in the equations of motion. Analogous to the Riemann tensor dependence on the Levi-Civita connection, we can write a torsion tensor as \cite{Hayashi:1979qx}
\begin{equation}
    \udt{T}{\sigma}{\mu\nu} :=2\udt{\Gamma}{\sigma}{[\nu\mu]}\,,
\end{equation}
where square brackets denote the anti-symmetry operator, and where torsion is the result of anti-symmetry \cite{Aldrovandi:2013wha}. The torsion tensor is covariant under both diffeomorphisms and local Lorentz transformations. The torsion tensor can be contracted in such a way to produce the torsion scalar \cite{Krssak:2018ywd,Cai:2015emx,Aldrovandi:2013wha,Bahamonde:2021gfp}
\begin{equation}\label{eq:torsion_scalar_def}
    T:=\frac{1}{4}\udt{T}{\alpha}{\mu\nu}\dut{T}{\alpha}{\mu\nu} + \frac{1}{2}\udt{T}{\alpha}{\mu\nu}\udt{T}{\nu\mu}{\alpha} - \udt{T}{\alpha}{\mu\alpha}\udt{T}{\beta\mu}{\beta}\,,
\end{equation}
which is the result of demanding that an action based only on $T$ gives the same equations of motion as the regular Einstein-Hilbert action with $\lc{R}$ (up to a boundary term). Analogous to the curvature scalar being solely dependent on the Levi-Civita connection, the torsion scalar ultimately is dependent only on the teleparallel connection.

The curvature scalar will organically vanish when calculated using the teleparallel connection since this is a measure of curvature and the teleparallel connection has none, i.e. $R\equiv 0$ (where we emphasize that $R = R(\udt{\Gamma}{\sigma}{\mu\nu})$ and $\lc{R}=\lc{R}(\udt{\lc{\Gamma}}{\sigma}{\mu\nu}) \neq 0$). In this background, we can write an expression relating the curvature and torsion scalars with each other such that \cite{Bahamonde:2015zma,Farrugia:2016qqe}
\begin{equation}\label{LC_TG_conn}
    R=\lc{R} + T - B = 0\,.
\end{equation}
where $B$ represents a total divergence term and is defined as
\begin{equation}\label{eq:boundary_term_def}
    B = \frac{2}{e}\partial_{\rho}\left(e\udut{T}{\mu}{\mu}{\rho}\right)\,,
\end{equation}
where $e=\det\left(\udt{e}{a}{\mu}\right)=\sqrt{-g}$ is the determinant of the tetrad. These relations guarantee the equivalence of the GR with TEGR, where TEGR has an action based on the torsion scalar analogous to the Einstein-Hilbert being based on the curvature scalar.

As in GR, we can consider extensions to the TEGR action where TEGR is complemented by additional terms which may embody possibly dynamical features of dark energy. In this work, we explore the generalization of the TEGR action to an arbitrary function of both the torsion scalar and the boundary term, which gives \cite{Bahamonde:2015zma,Capozziello:2018qcp,Bahamonde:2016grb,Paliathanasis:2017flf,Farrugia:2018gyz,Bahamonde:2016cul,Wright:2016ayu}
\begin{equation}\label{f_T_B_ext_Lagran}
    \mathcal{S}_{f(T,B)}^{} =  \frac{1}{2\kappa^2}\int \mathrm{d}^4 x\; e\,f(T,B) + \int \mathrm{d}^4 x\; e\mathcal{L}_{\text{m}}\,,
\end{equation}
where $\kappa^2=8\pi G$, and $\mathcal{L}_{\text{m}}$ is the matter Lagrangian in the Jordan frame. TEGR is recovered for the limit where $f(T,B) \rightarrow -T$. In this description the second and fourth order contributions to the equations of motion are decoupled with the respective torsion and boundary term scalars. In this context, there is a generalization of the fourth-order $f(\lc{R})$ gravity \cite{Sotiriou:2008rp,Capozziello:2011et} which now is recovered for the limit $f(T,B) \rightarrow f(-T+B) = f(\lc{R})$ (on comparison with Eq.~(\ref{LC_TG_conn})). Taking a variation with respect to the tetrad the produces the field equations \cite{Bahamonde:2015zma,Farrugia:2018gyz}
\begin{eqnarray}
    \udt{W}{\nu}{\lambda} := 2\delta_{\nu}^{\lambda}\mathring{\Box}f_{B}-2\mathring{\nabla}^{\lambda}\mathring{\nabla}_{\nu}f_{B}+Bf_{B}\delta_{\nu}^{\lambda}+4\Big[(\partial_{\mu}f_{B})+(\partial_{\mu}f_{T})\Big]S_{\nu}{}^{\mu\lambda}& &\nonumber \\
    +4e^{-1}e^{A}{}_{\nu}\partial_{\mu}(eS_{A}{}^{\mu\lambda})f_{T}-4f_{T}T^{\sigma}{}_{\mu\nu}S_{\sigma}{}^{\lambda\mu}-f\delta_{\nu}^{\lambda} & =& 2\kappa^{2}\udt{\Theta}{\nu}{\lambda}\,,\label{field_equations}
\end{eqnarray}
where subscripts denote derivatives, and $\udt{\Theta}{\nu}{\lambda}$ is the regular energy-momentum tensor for matter.

At this juncture, it is important to point out that the tetrad and spin connection are independent variables in the theory and so produce independent field equations. The tetrad produces the ten metrical field equations while the spin connection produces the six local Lorentz field equations. However, TG has a very useful property whereby the tetrad variation is connected to the spin connection field equations through $\dut{W}{a}{\mu} = \delta \mathcal{S}_{\mathcal{F}(T)}^{}/ \delta \udt{e}{A}{\mu}$ \cite{Hohmann:2021fpr}. The property states that the tetrad and spin connection field equations are respectively the result of the symmetric and anti-symmetric operator on $W_{\mu\nu}$. Due to the symmetry of the energy-momentum tensor, this means that these field equations can be written as \cite{Bahamonde:2021gfp}
\begin{equation}
    W_{(\mu\nu)} = \kappa^2 \Theta_{\mu\nu}\,, \quad \text{and} \quad W_{[\mu\nu]} = 0\,,
\end{equation}
which holds for any formalism based on the teleparallel connection. For particular choices of tetrads, the anti-symmetric field equations can be satisfied for vanishing spin connection components. This is called the Weitzenb\"{o}ck gauge and can be helpful for making calculations.

In this work, we consider a flat Friedmann–Lema\^{i}tre–Robertson–Walker (FLRW) cosmology where the tetrad can be written as \cite{Escamilla-Rivera:2019ulu}
\begin{equation}\label{flrw_tetrad}
    \udt{e}{A}{\mu}=\textrm{diag}(1,a(t),a(t),a(t))\,,
\end{equation}
where $a(t)$ is the scale factor, and which produces the flat homogeneous and isotropic metric
\begin{equation}
    ds^2=dt^2-a(t)^2(dx^2+dy^2+dz^2)\,,
\end{equation}
through Eq.~(\ref{metric_tetrad_rel}). This diagonal form of the tetrad is compatible with the Weitzenb\"{o}ck gauge, i.e. $\udt{\omega}{A}{B\mu}=0$ \cite{Krssak:2015oua,Tamanini:2012hg}. Using the definitions in Eqs.~(\ref{eq:torsion_scalar_def},\ref{eq:boundary_term_def}), we can write
\begin{equation}
    T =-6H^2\,,\quad B =-6(3H^2+\dot{H})\,,
\end{equation}
which together reproduce the regular curvature-based curvature scalar $\lc{R}=-T+B = -6(\dot{H} + 2H^2)$. The Friedmann equations can also be determined giving
\begin{eqnarray}
    3H(\dot{f}_{B}-2Hf_{T})+\frac{1}{2}(Bf_{B}-f) &=& \kappa^{2}\rho\,,\label{eq:Friedmann1}\\
    -\ddot{f}_{B}+2f_{T}\dot{H}+2H(3Hf_{T}+\dot{f}_{T})+\frac{1}{2}(f - Bf_{B}) &=& \kappa^{2}P\,,\label{eq:Friedmann2}
\end{eqnarray}
where overdots refer to derivatives with respect to cosmic time $t$, and where $\rho$ and $P$ respectively represent the energy density and pressure of matter.

For background cosmology, we can rewrite these equations with the extension to the TEGR Lagrangian term as an effective fluid which embodies the additional terms that are sourced from the extra torsion contributions. We first consider the mapping $f(T,B)\rightarrow-T + \mathcal{F}(T,B)$. Then, the equations of motion in Eqs.~(\ref{eq:Friedmann1},\ref{eq:Friedmann2}) take the form
\begin{eqnarray}
    3H^2 &=& \kappa^2 \left(\rho+\rho_{\text{eff}}\right)\,,\\
    3H^2 + 2\dot{H} &=& -\kappa^2\left(P+P_{\text{eff}}\right)\,,
\end{eqnarray}
where the effective fluid is defined through
\begin{eqnarray}
    \kappa^2 \rho_{\text{eff}} &:=& 3H^2\left(3\mathcal{F}_B + 2\mathcal{F}_T\right) - 3H\dot{\mathcal{F}}_B + 3\dot{H}\mathcal{F}_B + \frac{1}{2}\mathcal{F}\,, \label{eq:friedmann_mod}\\
    \kappa^2 P_{\text{eff}} &:=& -\frac{1}{2}\mathcal{F}-\left(3H^2+\dot{H}\right)\left(3\mathcal{F}_B + 2\mathcal{F}_T\right)-2H\dot{\mathcal{F}}_T+\ddot{\mathcal{F}}_B\,.
\end{eqnarray}

This effective fluid for the $\mathcal{F}(T,B)$ gravity extension satisfies the fluid equation \cite{Bahamonde:2016grb}
\begin{equation}
    \dot{\rho}_{\text{eff}}+3H\left(\rho_{\text{eff}}+P_{\text{eff}}\right) = 0\,,
\end{equation}
and can be used to define an effective equation of state (EoS)
\begin{eqnarray}\label{EoS_func}
    \omega_{\text{eff}} &:=& \frac{P_{\text{eff}}}{\rho_{\text{eff}}}\\ & = & -1+\frac{\ddot{\mathcal{F}}_B-3H\dot{\mathcal{F}}_B-2\dot{H}\mathcal{F}_T-2H\dot{\mathcal{F}}_T}{3H^2\left(3\mathcal{F}_B+2\mathcal{F}_T\right)-3H\dot{\mathcal{F}}_B+3\dot{H}\mathcal{F}_B-\frac{1}{2}\mathcal{F}}\,. 
\end{eqnarray}
In the $\Lambda$CDM limit, as expected this EoS parameter limits to $\omega_{\text{eff}} \rightarrow -1$.

%%%%%%%%%%%%%%%%%%%%%%%%%%%%%%%%%%%%%%%%%%%%%%%%
%%%%%%%%%%%%%%%%%%%%%%%%%%%%%%%%%%%%%%%%%%%%%%%%

\section{\label{sec:obs_data}Observational Data}

We here present and briefly discuss the observational data sets which will be considered in the below analyses. For our baseline data set, we consider Hubble expansion data along with an SNe Ia compilation data set.

\begin{itemize}
    \item For the Hubble parameter data, we adopt thirty--one cosmic chronometer data points \cite{2014RAA....14.1221Z,Jimenez:2003iv,Moresco:2016mzx,Simon:2004tf,2012JCAP...08..006M,2010JCAP...02..008S,Moresco:2015cya}. It is well-known that such a technique enables us to directly derive information about the Hubble function at several redshifts, up to $z\lesssim2$. Since the adopted CC data is primarily based on measurements of the age difference between two passively--evolving galaxies that formed at the same time but are separated by a small redshift interval, CC were found to be more reliable than any other method based on an absolute age determination for galaxies \cite{Jimenez:2001gg}. We should also remark that our CC data set is independent of the Cepheid distance scale and also considered to be model independent, although CC data points are known to be dependent on the modelling of stellar ages, which is based on robust stellar population synthesis techniques (see, for instance, Refs. \cite{Gomez-Valent:2018hwc,Lopez-Corredoira:2017zfl,Lopez-Corredoira:2018tmn,Verde:2014qea,2012JCAP...08..006M,Moresco:2016mzx} for analyses related to CC systematics).

\item We shall also be considering the Pantheon SNe Ia  data compilation consisting of 1048 SNe Ia relative luminosity distance measurements spanning the redshift range of $0.01<z<2.3$ \cite{Scolnic:2017caz}. Henceforth, we will be denoting the Pantheon SNe Ia compilation by SN. We should remark that since the apparent magnitude of each SNe Ia needs to be calibrated via an arbitrary fiducial absolute magnitude $M$, we will be considering $M$ as a nuisance parameter in our Markov chain Monte Carlo (MCMC) analyses\footnote{We perform out analysis using the \textit{emcee} package available at Ref.~\cite{2013PASP..125..306F}.}. 

\item Furthermore, we shall also be considering a joint baryon acoustic oscillation (BAO) data set. The considered BAO data set is composed of the SDSS Main Galaxy Sample measurement at $z_{\mathrm{eff}}=0.15$ \cite{Ross:2014qpa}, the six--degree Field Galaxy Survey measurement at $z_{\mathrm{eff}}=0.106$ \cite{2011MNRAS.416.3017B}, and the BOSS DR11 quasar Lyman--$\alpha$ measurement at $z_{\mathrm{eff}}=2.4$ \cite{Bourboux:2017cbm}. We further consider the angular diameter distances and $H(z)$ measurements of SDSS--IV eBOSS DR14 quasar survey at $z_{\mathrm{eff}}=\{0.98,\,1.23,\,1.52,\,1.94\}$ \cite{Zhao:2018gvb}, along with the SDSS--III BOSS DR12 consensus BAO measurements of the Hubble parameter and the corresponding comoving angular diameter distances at $z_{\mathrm{eff}}=\{0.38,\,0.51,\,0.61\}$ \cite{Alam:2016hwk}, where in these two BAO data sets we consider the full covariance matrix in our MCMC analyses. For each set of parameters in the MCMC analysis, we computed the comoving sound horizon $r_s(z)$ at the end of the baryon drag epoch at redshift $z_d\approx1059.94$ \cite{Planck:2018vyg} via an accurate expression which has been inferred from genetic algorithms \cite{Aizpuru:2021vhd}, in which we have adopted a baryon content of $\Omega_{b,0} = 0.02166\,h^{-2}$ with $h=H_0/100$.
\end{itemize}

We shall also be analysing the impact of $H_0$ prior values on our inferred $f(T,B)$ model parameter constraints. We will be considering the latest SH0ES local estimate \cite{Riess:2021jrx} of $H_0= 73.30 \pm 1.04 \,{\rm km\, s}^{-1} {\rm Mpc}^{-1}$ (R21) based on SN in the Hubble flow, and the measurement using the tip of the red giant branch (TRGB) as a standard candle \cite{Freedman:2021ahq} with $H_0=69.8 \pm 1.7 \,{\rm km\, s}^{-1} {\rm Mpc}^{-1}$ (F21). 

%%%%%%%%%%%%%%%%%%%%%%%%%%%%%%%%%%%%%%%%%%%%%%%%
%%%%%%%%%%%%%%%%%%%%%%%%%%%%%%%%%%%%%%%%%%%%%%%%

\section{\label{sec:models}Constraints of \texorpdfstring{$f(T,B)$}{fTB} Cosmological Models}

In this work, we consider three $f(T,B)$ models together with two recent cosmological model independent $H_0$ priors which will test the consistency of these models with the priors. There exists several measurements of $H_0$ in literature as studies have shown how well they can reproduce our cosmological history. We examine the effect that these priors have on the model parameter values for the data sets mentioned in Sec.~\ref{sec:obs_data}. In this paper, we implement an MCMC algorithm using a combination of two choices. The first being the data sets chosen, i.e CC+SN or CC+SN+BAO and the second being the selection from having no prior, or one of the $H_0$ prior values (either R21 or F21).

In addition, we make a comparison of each $f_i$CDM model against the $\Lambda$CDM model by calculating the Akaike information criteria (AIC) \cite{Akaike:1974} and the Bayesian information criteria (BIC) \cite{10.1214/aos/1176344136}. This allows us to evaluate which model best fits the observational data. The AIC is defined as 
\begin{equation}\label{eq:AIC}
    \text{AIC} = -2 \ln L_{\text{max}} + 2n \,,
\end{equation}
where $L_{\text{max}}$ is the maximum value of the likelihood of each model, which includes likelihood of the data set and prior combination, and $n$ is the number of parameters involved in the MCMC algorithm. The AIC reward goodness of fit through the maximum likelihood, however, the additional term of the AIC serves as a penalty for models which have a large number of parameters. The BIC is defined as 
\begin{equation} \label{eq:BIC}
    \text{BIC} = -2\ln L_{\text{max}} + n \ln m \,,
\end{equation}
where $m$ is the number of data points used in the fit. From Eq.~\eqref{eq:BIC}, it is evident that the penalty for BIC is higher than that of AIC. Moreover, the involvement of the number of data points in the BIC criterion, makes it a potentially better measure of performance than the AIC criterion. In general, the model with lower AIC and BIC values corresponds to the model that best supports the data. 

For our model comparison, we calculate the $\Delta$AIC and $\Delta$BIC, where we compare each model data set and prior combination with the corresponding $\Lambda$CDM (See Appendix \ref{sec:app}). The difference between the two models can be defined as $\Delta \text{AIC} = \Delta \chi^2_{\text{min}} + 2 \Delta n$ and $\Delta \text{BIC} = \Delta \chi^2_{\text{min}} + \Delta n \ln m$ for AIC and BIC, respectively. In this context, the small values of $\Delta$AIC and $\Delta$BIC correspond to the model with the chosen data set and prior combination being closer to $\Lambda$CDM.

%%%%%%%%%%%%%%%%%%%%%%%%%%%%%%%%%%%%%%%%%%%%%%%%
%%%%%%%%%%%%%%%%%%%%%%%%%%%%%%%%%%%%%%%%%%%%%%%%

\subsection{Power Law Model -- \texorpdfstring{$f_1(T,B)$}{}CDM Model}

The power law model was first considered by Bengochea and Ferraro in \cite{Bengochea:2008gz} because of its ability to reproduce the late-time cosmic acceleration in $f(T)$ gravity. The $f(T,B)$ model is given by
\begin{equation}\label{eq:PLM}
    \mathcal{F}_1(B) = \alpha_1 \left( -B\right)^{p_1}\,,
\end{equation}
where $\alpha_1$ and $p_1$ are constants. From the modified Friedmann equation~\ref{eq:Friedmann1}, we can obtain $\alpha_1$ by evaluating at current time
\begin{equation}
    \alpha_1 = \frac{6H_0^2 (\Omega_{m_0} + \Omega_{r_0} - 1)}{(p_1-1)(-B_0)^{p_1} + 6H_0\left(p_1(p_1-1)(-B_0)^{p_1-2} \dot{B}\vert_{t=t_0}\right)} \,,
\end{equation}
where $\Omega_{m_0}$ and $\Omega_{r_0}$ are the matter and radiation density parameters at current times, respectively. This expression of $\alpha_1$ in terms of $p_1$ and the other cosmological parameters is very helpful in that it reduces the number of free parameters of the system, making the analysis less intensive. Moreover, $B_0$ is the boundary term value at current times, i.e $B_0 = B\vert_{t=t_0}$ whilst $\dot{B}\vert_{t=t_0}$ is its derivative also evaluated at current times. Thus, $p_1$ is the only new model parameter for $f_1(T,B)$CDM model.

Thus, the Friedmann equation in Eq.~\ref{eq:Friedmann1} for this model can be written as
\begin{align}
    H'' &= \frac{-1}{36\alpha_1(1+z)^2H^3 p_1(p_1-1)(-B)^{p_1-2}} \nonumber \\
    & + \Big[ 6H_0^2 \Big(\Omega_{m_0}(1+z)^3 + \Omega_{r_0}(1+z)^4 \Big) - 6H_0^2 + \alpha_1(1-p_1)(-B)^{p_1} - 216 \alpha_1 p_1 (p_1-1) H^3H'(1+z)(-B)^{p_1-2}\Big] \\ 
    &-\frac{H'}{1+z} - \frac{H'^2}{H} \nonumber\,,
\end{align}
where we have transformed to redshift space, and where primes refer to derivatives with respect to redshift $z$. The above model reduces to the $\Lambda$CDM model when $p_1 =0$. When $p_1 = 1$, we obtain the GR limit, which gives an upper bound, $p_1< 1$, such that we have an accelerating Universe.

\begin{table}
\small
    \centering
    \caption{Results for the Power Law Model (\ref{eq:PLM}). First column contains the data sets used to constrain parameters together with their $H_0$ priors. Second to fourth column include constrained output parameters for $\{H_0, \Omega_{m,0}, p_1\}$ respectively. Fifth column gives the nuisance parameter, $M$. Sixth column contains $\chi^2_{\textrm{min}}$ whilst seventh up to tenth column includes the statistical indicators and their difference with respect to $\Lambda$CDM model.}
    \label{tab:PLM}
    \begin{tabular}{>{\centering}m{0.18\textwidth}>{\centering}m{0.1\textwidth}>{\centering}m{0.12\textwidth}>{\centering}m{0.12\textwidth}>{\centering}m{0.12\textwidth}>{\centering}m{0.06\textwidth}>{\centering}m{0.06\textwidth}>{\centering}m{0.06\textwidth}>{\centering}m{0.06\textwidth}>{\centering\arraybackslash}m{0.06\textwidth}}
        \hline
		\rule{0pt}{1.2 \baselineskip}\rule{0pt}{1.2 \baselineskip}Data Sets & $H_0$[\si{\km/ \s / \mega \parsec}] & $\Omega_{m,0}$ & $p_1$ & $M$ & $\chi^2_\mathrm{min}$& AIC & BIC & $\Delta$AIC & $\Delta$BIC \\ [4pt]
		\hline
		\rule{0pt}{1.2 \baselineskip}CC + SN & $68.5^{+2.1}_{-2.2}$ & $0.281^{+0.027}_{-0.023}$ & $0.140^{+0.090}_{-0.109}$ & $-19.381^{+0.063}_{-0.064}$ & 1040.94 & 1048.94 & 1068.88 & $1.21$ & $6.19$ \\[4pt] 
		$ \mathrm{CC + SN + R21}$ & $72.0\pm 1.1$ & $0.269^{+0.021}_{-0.016}$ & $0.080^{+0.108}_{-0.068}$ & $-19.27^{+0.19}_{-0.22}$ & 1040.50 & 1053.50 & 1076.10 & $1.04$ & $6.02$ \\ [4pt]
		$\mathrm{CC+ SN + F21}$ & $69.1^{+1.5}_{-1.4}$ & $0.279^{+0.025}_{-0.022}$ & $0.143^{+0.079}_{-0.113}$ & $-19.362^{+0.075}_{-0.078}$ & 1040.98 & 1048.98 & 1068.92 & $1.10$ & $6.08$  \\ [4pt]
		\hline
		\rule{0pt}{1.2 \baselineskip}CC + SN + BAO & $67.5\pm 1.4$ & $0.305\pm 0.017$ & $0.058\pm 0.053$ & $-19.396^{+0.090}_{-0.136}$ & 1047.97 & 1055.97 & 1075.97 & $-0.04$  & $4.96$  \\ [4pt]
		$\mathrm{CC + SN + BAO + R21}$ & $70.32^{+0.95}_{-0.94}$ & $0.314\pm 0.016$ & $ 0.012^{+0.027}_{-0.0094} $ & $-19.327^{+0.029}_{-0.034}$ & 1062.23 & 1070.23 & 1090.24 & 0.79  & 5.79 \\ [4pt]
		$\mathrm{CC+ SN + BAO + F21}$ & $68.1\pm 1.2$ & $0.308\pm 0.017$ & $0.046^{+0.045}_{-0.042}$ & $-19.35^{+0.69}_{-0.78}$ & 1048.81 & 1056.81 & 1076.81 & 0.13 & 5.12 \\ [4pt]
		\hline
    \end{tabular}
\end{table}

\begin{figure}[H]
    \centering
    \includegraphics[scale = 0.35]{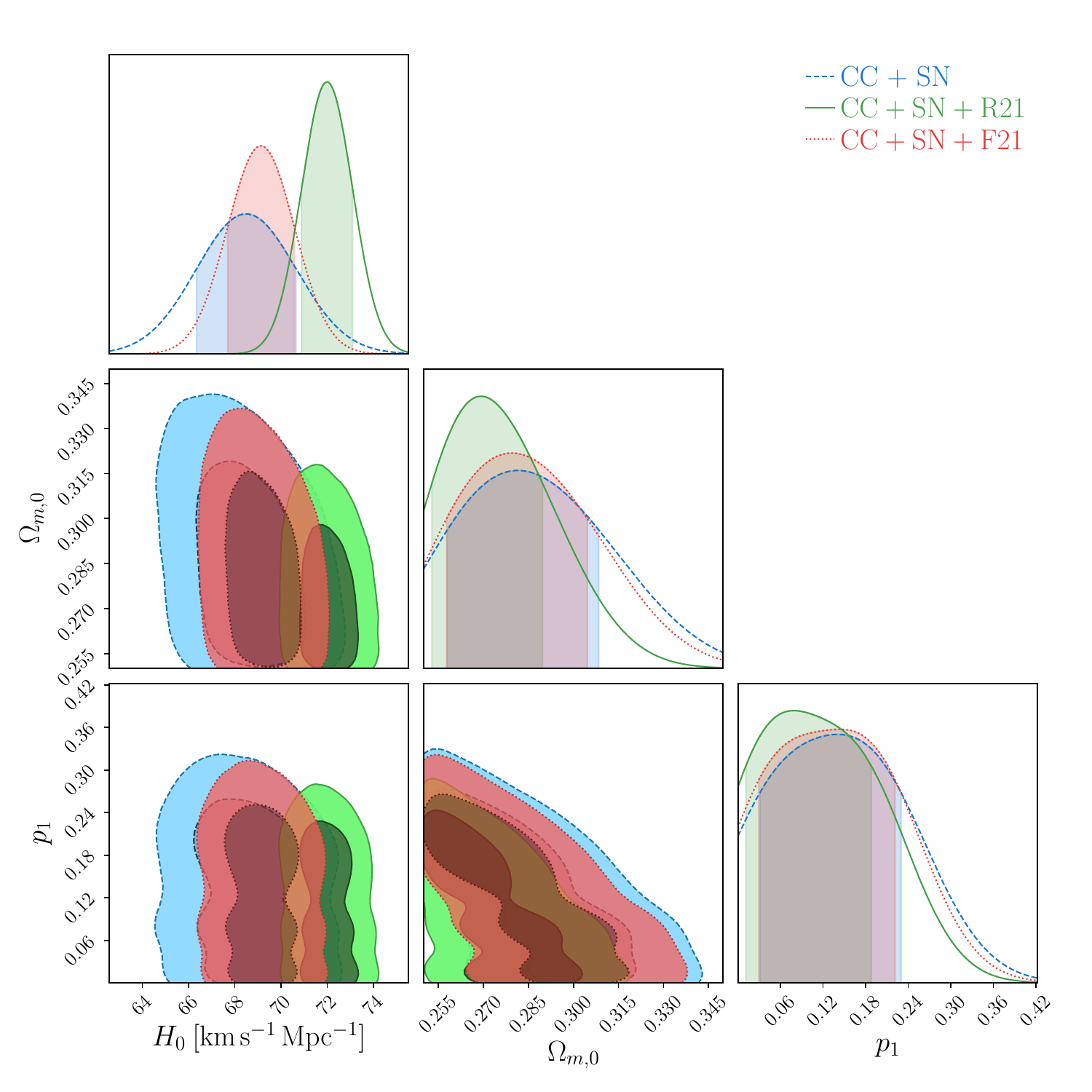}
     \includegraphics[scale = 0.35]{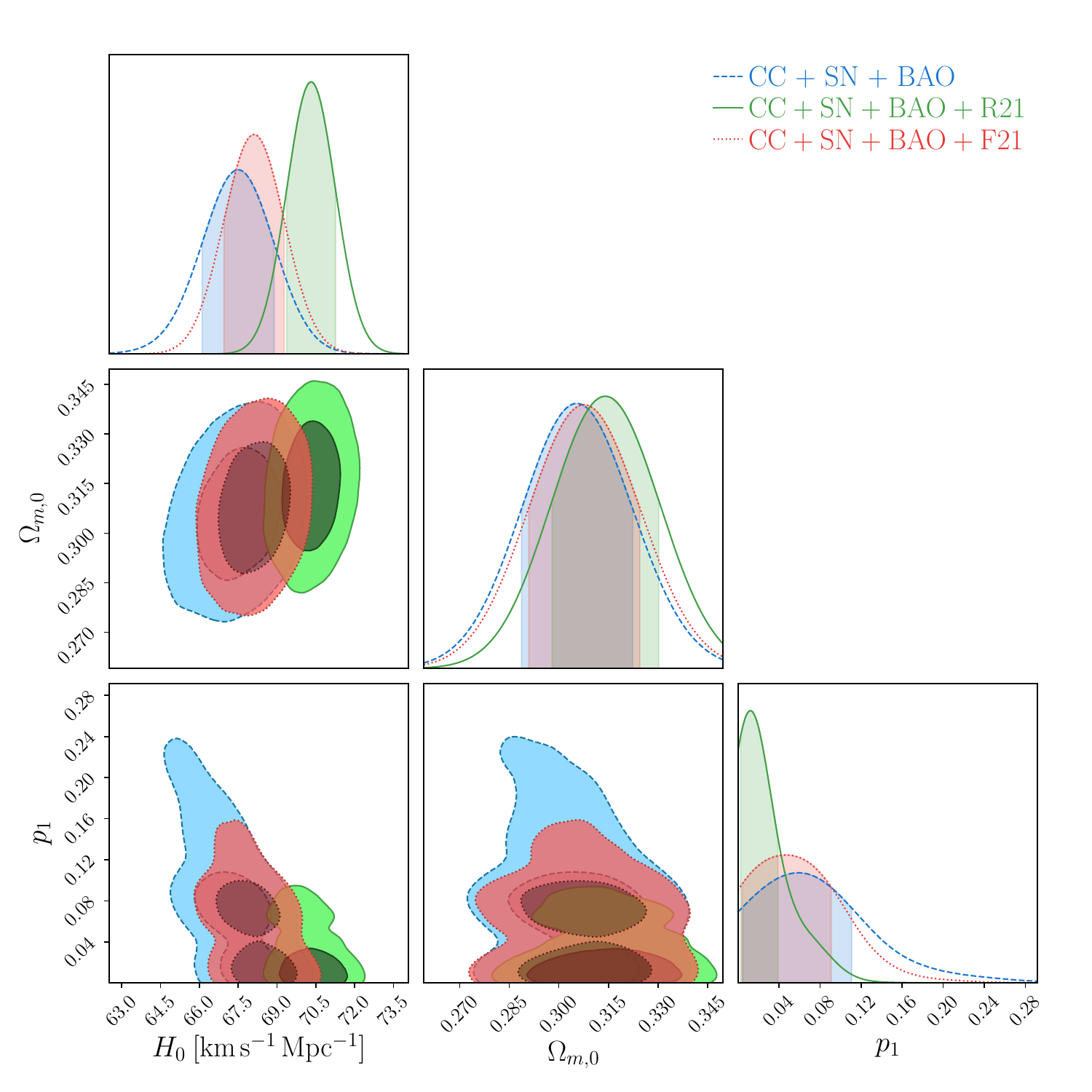}
    \caption{Confidence level (C.L) for the Power Law model Eq.~(\ref{eq:PLM}). The regions in color denotes: CC+SN sample (blue), CC+SN+R21 (green) and CC+SN+F21 (red). The priors R21 and F21 indicates the values quoted in Sec.~\ref{sec:obs_data}.}
    \label{fig:M1_CL}
\end{figure}

For this model, we use the MCMC algorithm to plot the posteriors and the confidence regions for CC+SN and CC+SN+BAO data sets shown in Fig.~\ref{fig:M1_CL}. In addition, we also plot the results for each prior of $H_0$ mentioned in Sec.~\ref{sec:obs_data}. The most apparent feature of the results is the impact that these priors have on $H_0$ values as one immediately notices that these priors tend to increase the Hubble constant value when compared to the no prior situation. Similarly, although to a lesser extent, $\Omega_{m_0}$ has a lower value when priors are added to the chosen data sets, which is to be expected since the priors shift the value of $H_0$. There is also some anti-correlation between the $\Omega_{m,0}$ and $p_1$ parameters for the CC+SN data sets. An interesting feature of this model is that for all cases of the CC+SN+BAO data sets there are two possible valid sets for $p_1$ for $H_0$. Correspondingly, in a Hubble diagram of $H(z)$ versus redshift, $z$, the two values of the parameter $p_1$ are changing the tilt of the graph such that an $H_0$ within the same range is created. Similarly, this feature is also attributed to the two parameters $\Omega_{m,0}$ and $p_1$, where for a set of value of $\Omega_{m,0}$, $p_1$ can have two valid sets.

The exact values for the model parameters obtained from the MCMC runs are found in Table.\ref{tab:PLM}, together with the values of $\chi^2_{\text{min}}$, AIC, BIC and $\Delta$AIC and $\Delta$BIC. The first column confirms the fact that the priors tend to raise the value of $H_0$ and, in fact the highest value of the Hubble constant is obtained for the CC+SN with the R21 prior ($H_0 = 72.0\pm 1.1 \,{\rm km\, s}^{-1} {\rm Mpc}^{-1}$). This, however, is to be expected as the R21 prior has a higher value than the F21 prior and thus, it tends to shift the value of $H_0$ to a higher value. Consequently, the $\Omega_{m,0}$ parameter for CC+SN+R21 has the lowest value, which agrees with having a large value of the Hubble constant. On the other hand, the minimum value of $H_0$ is reached for the CC+SN+BAO data set with no prior ($H_0 = 67.5\pm 1.4 \,{\rm km\, s}^{-1} {\rm Mpc}^{-1}$), which is also foreseeable since the BAO data is dependent on the early Universe. In turn, this raises the value of $\Omega_{m,0}$ to the highest value.

For all cases, the $p_1$ parameter shows slight deviation from the $\Lambda$CDM model as the values for such parameter fall within 2$\sigma$ of the standard model of cosmology. An interesting feature in this model is that the $\Delta$AIC for CC+SN+BAO is slightly negative, meaning that the maximum likelihood of this model with the CC+SN+BAO with no prior is marginally higher than that of the $\Lambda$CDM. In consequence, one can say that this model with this particular data set is slightly preferred over the $\Lambda$CDM. Coincidentally, the BIC for such data set also has the lowest value.

%%%%%%%%%%%%%%%%%%%%%%%%%%%%%%%%%%%%%%%%%%%%%%%%
%%%%%%%%%%%%%%%%%%%%%%%%%%%%%%%%%%%%%%%%%%%%%%%%
\subsection{Exponential Model -- \texorpdfstring{$f_2(T,B)$}{} Model}

The exponential model~\cite{Linder:2010py} can be described by 
\begin{equation}\label{eq:EM}
    \mathcal{F}_2(B) = \alpha_2 \,\text{Exp}\left[-p_2 \sqrt{\frac{B}{B_0}} \right] \,,
\end{equation}
where $\alpha_2$ and $p_2$ are constants. These constants can be related by evaluating the Friedmann equation at current time such that
\begin{equation}
    \alpha_2 = \frac{e^{p_2} 6H_0^2 (\Omega_{m_o} + \Omega_{r_0} -1)}{-1 - \frac{p_2}{2} + \frac{6H_0 p_2}{4B_0^2} (1+ p_2) \dot{B}\vert_{t=t_0} } \,,
\end{equation}
making $p_2$ the new model parameter for this model. Thus, the Friedmann equation can then be written as 
\begin{align}
    H'' & = \frac{-BB_0 \text{Exp}\left[p_2\sqrt{\frac{B}{B_0}}\right]}{9\alpha_2p_2 (1+z)^2H^3 \left(p_2 \sqrt{\frac{B_0}{B}} \right)} \nonumber \\
    &\Bigg[ 6H_0^2 \Big(\Omega_{m_0}(1+z)^3 + \Omega_{r_0}(1+z)^4 \Big) - 6H_0^2  + \alpha_2\text{Exp}\left[p_2\sqrt{\frac{B}{B_0}} \right] \left(1 - \frac{54 p_2 (1+z) H^3 H'}{BB_0}\left( p_2 + \sqrt{\frac{B_0}{B}}\right) \right) \Bigg]\\
    &-\frac{H'}{1+z} - \frac{H'^2}{H} \nonumber \,.
\end{align}
which reduces to $\Lambda$CDM when $p_2 = 0$

\begin{figure}[H]
    \centering
    \includegraphics[scale = 0.35]{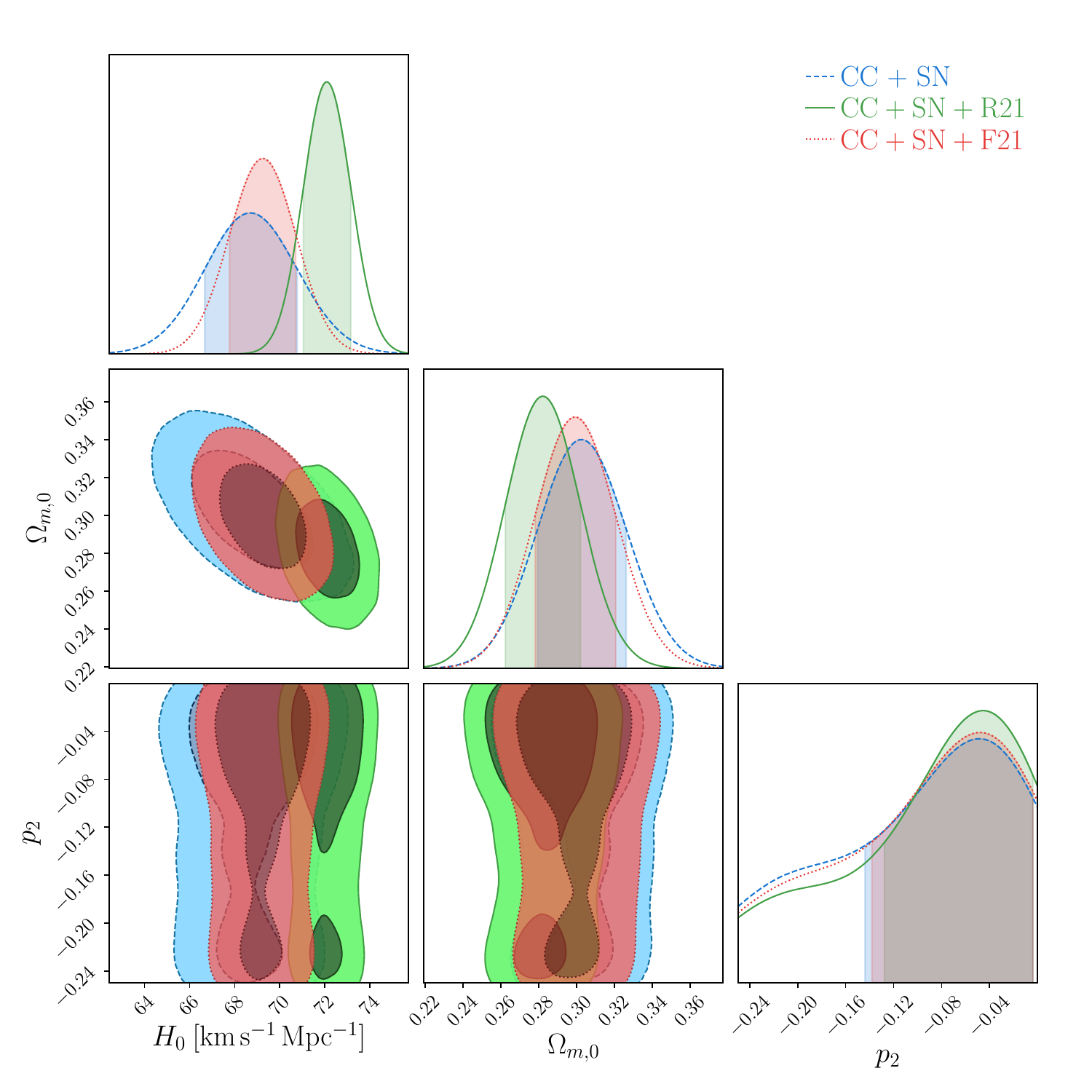}
     \includegraphics[scale = 0.35]{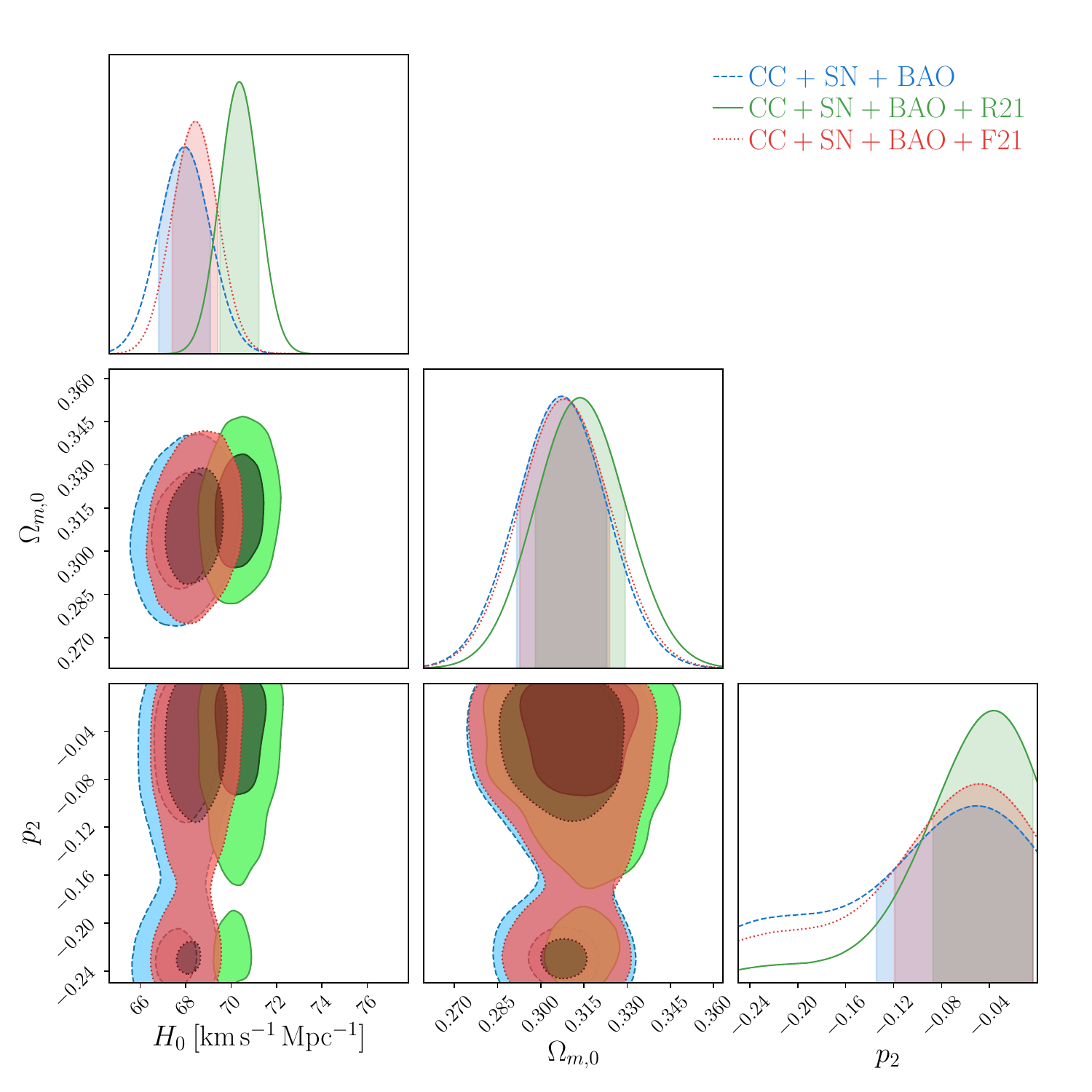}
    \caption{C.L for the Exponential model Eq.~(\ref{eq:EM}). The regions in color denotes: CC+SN sample (blue), CC+SN+R21 (green) and CC+SN+F21 (red). The priors R21 and F21 indicates the values quoted in Sec.~\ref{sec:obs_data}.}
    \label{fig:M2_CL}
\end{figure}

The confidence regions together with the posteriors are shown in Fig.~\ref{fig:M2_CL} for all data sets. Similar to the power law model, the priors tend to shift the value of the Hubble constant to a higher value compared to when no priors are implemented, whilst $\Omega_{m,0}$, is shifted towards the lower values. The highest value of $H_0$ is in agreement with the previous model and is also reached for the CC+SN with the R21 prior ($H_0 = 72.1^{+1.1}_{-1.0} \,{\rm km\, s}^{-1} {\rm Mpc}^{-1}$ ). In addition, the value of $\Omega_{m,0}$ is the lowest in this case at a value of $\Omega_{m,0} = 0.307 \pm 0.016$. Indeed, the anti-correlation feature between the parameters $H_0$ and $\Omega_{m,0}$ is more evident in this model especially for the CC+SN data sets. On the other hand, the lowest value, is again obtained for the CC+SN+BAO ($H_0 = 68.0^{+1.1}_{-1.2} \,{\rm km\, s}^{-1} {\rm Mpc}^{-1}$). One downside of this analysis is that given the higher order nature of the model, the ensuing Friedmann equations turned out to take an exceedingly long time to solve numerically outside of the indicated parameter range for $p_2$. Saying that, a clear Gaussian can be observed for the posterior probability plot of this parameter.

The $p_2$ values in this model are even lower than $f_1$CDM and closer to zero. However, unlike $f_1$CDM, the values are within 1$\sigma$ of $\Lambda$CDM, and thus, this model seems to be closer to the $\Lambda$CDM than the power law model. The statistical indicators, however, are very similar to the previous model and again in this case the smallest values obtained for $\Delta$AIC and $\Delta$BC are for the data set CC+SN+BAO with no priors. Despite that, the values for such statistical indicators for the CC+SN+BAO with the F21 prior are very similar to the CC+SN+BAO with no prior and in fact, the values of the Hubble constant and the density parameter of both these data sets are very close to each other. 

\begin{table}
\small
    \centering
    \caption{Results for the Exponential Model (\ref{eq:EM}). First column contains the data sets used to constrain parameters together with their $H_0$ priors. Second to fourth column include constrained output parameters for $\{H_0, \Omega_{m,0}, p_2\}$ respectively. Fifth column gives the nuisance parameter, $M$. Sixth column contains $\chi^2_{\textrm{min}}$ whilst seventh up to tenth column includes the statistical indicators and their difference with respect to $\Lambda$CDM model.}
    \label{tab:EM}
    \begin{tabular}{>{\centering}m{0.18\textwidth}>{\centering}m{0.1\textwidth}>{\centering}m{0.12\textwidth}>{\centering}m{0.09\textwidth}>{\centering}m{0.14\textwidth}>{\centering}m{0.08\textwidth}>{\centering}m{0.06\textwidth}>{\centering}m{0.06\textwidth}>{\centering}m{0.06\textwidth}>{\centering\arraybackslash}m{0.06\textwidth}}
        \hline
		\rule{0pt}{1.2 \baselineskip}\rule{0pt}{1.2 \baselineskip}Data Sets & $H_0$[\si{\km/ \s / \mega \parsec}] & $\Omega_{m,0}$ & $p_2$ & $M$ & $\chi^2_\mathrm{min}$& AIC & BIC & $\Delta$AIC & $\Delta$BIC \\ [4pt]
		\hline
		\rule{0pt}{1.2 \baselineskip}CC + SN & $68.7^{+2.1}_{-2.0}$ & $0.302^{+0.024}_{-0.023}$ & $-0.049^{+0.045}_{-0.095}$ & $-19.398^{+0.115}_{-0.051}$ & 1040.70 & 1048.70 & 1068.64 & $0.96$ & $5.95$ \\[4pt] 
		$ \mathrm{CC + SN + R21}$ & $72.1^{+1.1}_{-1.0}$ & $0.282\pm 0.020$ & $-0.045^{+0.041}_{-0.082}$ & $-19.288^{+0.030}_{-0.031}$ & 1045.69 & 1053.69 & 1073.63 & $1.23$ & $6.22$ \\ [4pt]
		$\mathrm{CC+ SN + F21}$ & $69.3\pm 1.5$ & $0.299\pm 0.021$ & $-0.048^{+0.044}_{-0.090}$ & $-19.368^{+0.043}_{-0.044}$ & 1040.89 & 1048.89 & 1068.83 & $1.01$ & $6.00$  \\ [4pt]
		\hline
		\rule{0pt}{1.2 \baselineskip}CC + SN + BAO &  $68.0^{+1.1}_{-1.2}$ & $0.307\pm 0.016$ & $-0.050^{+0.046}_{-0.084}$ & $-18.95^{+0.25}_{-0.00}$& 1048.31 & 1056.31 & 1076.31 & 0.30 & 5.30  \\ [4pt]
		$\mathrm{CC + SN + BAO + R21}$ &  $70.38^{+0.85}_{-0.87}$ & $0.314\pm 0.016$ & $-0.037^{+0.033}_{-0.051}$ & $-19.329^{+0.053}_{-0.052}$ & 1062.46 & 1070.46 & 1090.46 & 1.02 & 6.02 \\ [4pt]
		$\mathrm{CC+ SN + BAO + F21}$ & $68.4^{+1.1}_{-1.2}$ & $0.307^{+0.016}_{-0.014}$ & $-0.044^{+0.046}_{-0.071}$ & $-18.7^{+7.7}_{-0.0}$  & 1049.22 & 1057.23 & 1077.23 & 0.54 & 5.54 \\ [4pt]
		\hline
    \end{tabular}
\end{table}

%%%%%%%%%%%%%%%%%%%%%%%%%%%%%%%%%%%%%%%%%%%%%%%%
%%%%%%%%%%%%%%%%%%%%%%%%%%%%%%%%%%%%%%%%%%%%%%%%

\subsection{Logarithmic Model -- \texorpdfstring{$f_3(T,B)$}{} Model}

The logarithmic model~\cite{Bamba:2010wb} is given by 
\begin{equation}\label{eq:LogM}
    \mathcal{F}_3 = \alpha_3 \ln \left[ p_3 \frac{B_0}{B}\right] \,,
\end{equation}
where $\alpha_3$ and $p_3$ are constants. By evaluating the Friedmann equation at current times we get
\begin{equation}
\alpha_3 = \frac{6 H_0^2 \left( \Omega{m_0} + \Omega{r_0} -1 \right)}{\frac{6H_0}{B_0^2}\dot{B}\vert_{t=t_0} -1 - \ln p_3} \,.
\end{equation}
The Friedmann equation, thus, can be written as 
\begin{align}
    H'' &= \frac{-B^2}{36 \alpha_3 (1+z)^2 H^3} \Bigg[ 6H_0^2 \Big(\Omega_{m_0}(1+z)^3 + \Omega_{r_0}(1+z)^4 \Big) - 6H_0^2 + \alpha_3 \ln \left[p_3 \frac{B_0}{B} \right] + \alpha_3 - \frac{216 \alpha_3}{B^2}H^3 H'(1+z) \Bigg] \\
     &-\frac{H'}{1+z} - \frac{H'^2}{H} \nonumber \,.
\end{align}
An interesting feature in this model is that no choice of parameter values can recover the $\Lambda$CDM. However, it is this lack of $\Lambda$CDM limit that provides an intriguing characteristic to this model as there are no expected values for $p_3$. 

\begin{figure}[H]
    \centering
    \includegraphics[scale = 0.35]{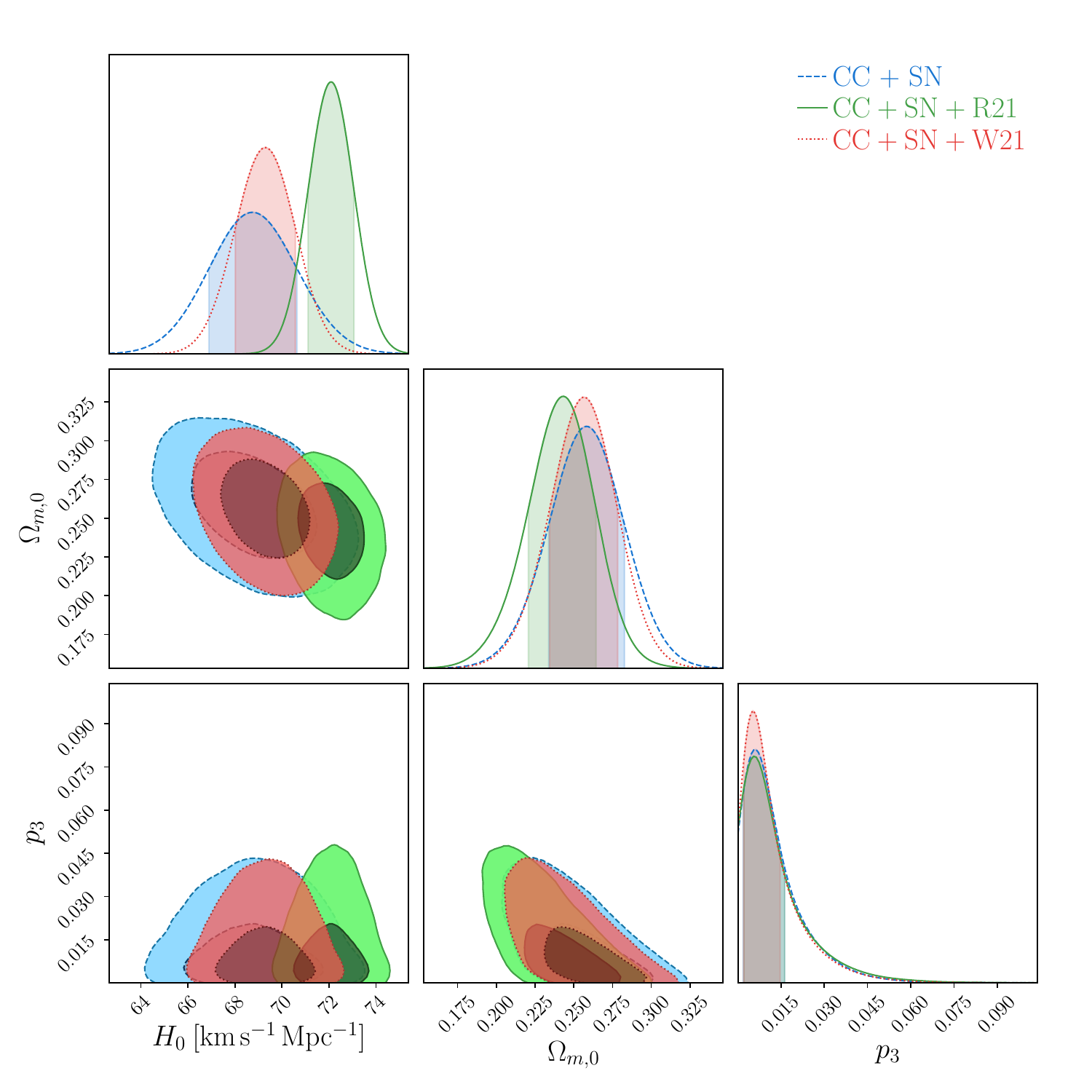}
     \includegraphics[scale = 0.35]{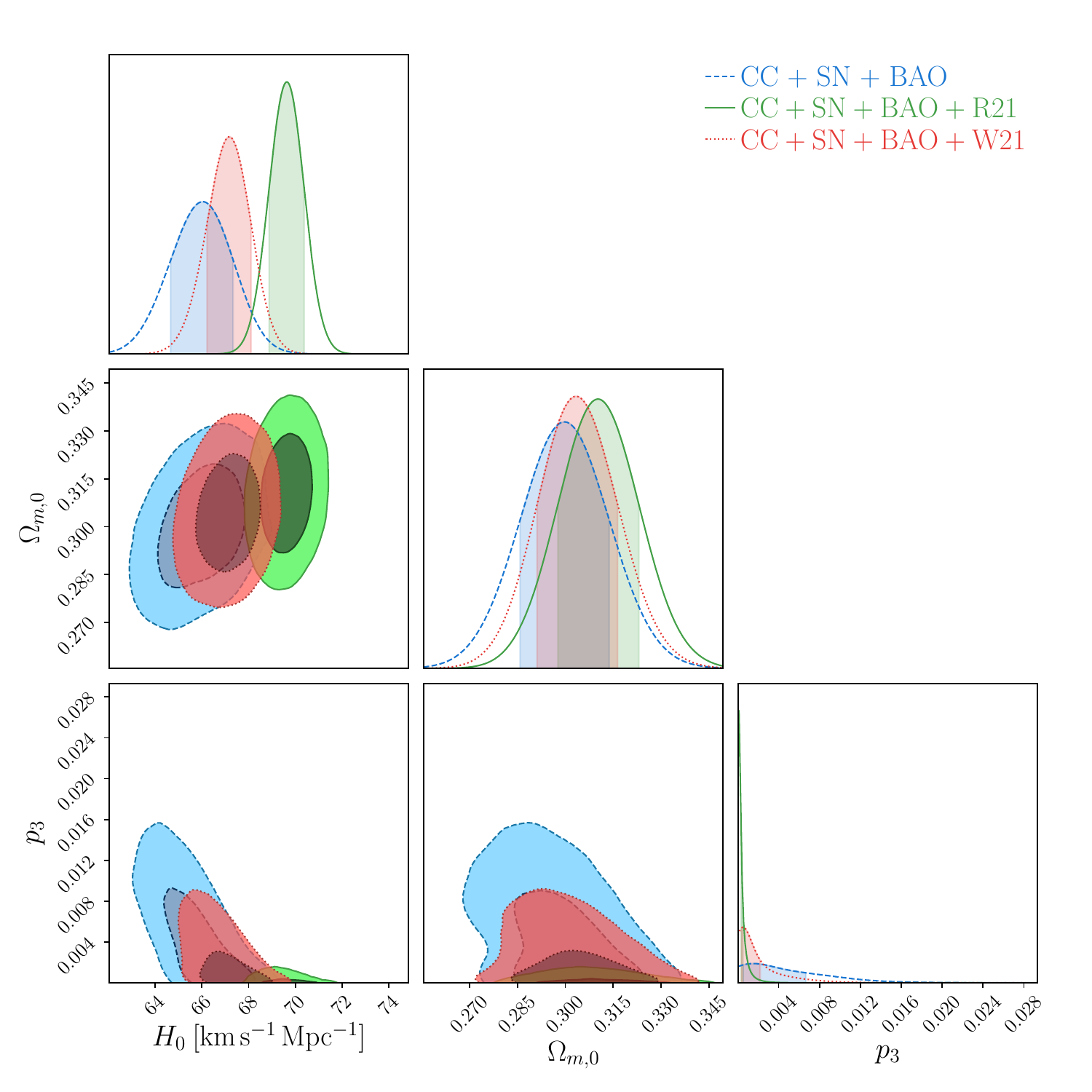}
    \caption{C.L for the Logarithmic model Eq.~(\ref{eq:LogM}). The regions in color denotes: CC+SN sample (blue), CC+SN+R21 (green) and CC+SN+F21 (red). The priors R21 and F21 indicates the values quoted in Sec.~\ref{sec:obs_data}.}
    \label{fig:M3_CL}
\end{figure}

Fig.\ref{fig:M3_CL} shows the posterior and confidence regions obtained through MCMC for all data sets. The precision results of such outputs are shown in Table. \ref{tab:LM}, in which one can notice that even though this model has no $\Lambda$CDM limit, there is no significant difference in the values of the Hubble constant compared to previous models. Indeed, the highest value obtained for $H_0$ is similar to that of the two preceding models and is again for the case of CC+SN with a R21 prior ($H_0 = 72.09^{+0.97}_{-0.98}\,{\rm km\, s}^{-1} {\rm Mpc}^{-1}$). In addition, the density parameter for CC+SN data prefers low values, irrespective of whether a prior was included or not. On the other hand, even though the lowest value of $H_0$ still appears for CC+SN+BAO, its magnitude turns out to be quite low in comparison to $f_1$CDM and $f_2$CDM, with $H_0 = 66.0^{+1.3}_{-1.4}{\rm km\, s}^{-1} {\rm Mpc}^{-1}$. Similar to the $f_2$CDM model, part of the potential values of the $p_3$ parameter were exceedingly difficult to solve for numerically due to the complexity of the higher order Friedmann equations. However, a clear Gaussian for the $p_3$ parameter can be observed.

Another noteworthy characteristic is that the values of $p_3$ for all choices of the data set is very small, close to zero. The statistical indicators also provide a slightly different scenario than preceding models, especially for the CC+SN data sets. In fact, the $\Delta$AIC and the $\Delta$BIC for such data sets have relatively higher values. Moreover, for CC+SN the values of $\Delta$AIC and the $\Delta$BIC are very similar regardless if a prior was included or not. The minimum value of $\Delta$AIC and the $\Delta$BIC also changes slightly, as in this case it is CC+SN+F21 which has the lowest value. Another crucial difference from previous models is the exceptionally high value of $\Delta$AIC and $\Delta$BIC for CC+SN+BAO+F21. These considerable differences from $\Lambda$CDM model challenges the suitability of this model for cosmology. 

\begin{table}
\small
    \centering
    \caption{Results for the Logarithmic Model (\ref{eq:LogM}). First column contains the data sets used to constrain parameters together with their $H_0$ priors. Second to fourth column include constrained output parameters for $\{H_0, \Omega_{m,0}, p_3\}$ respectively. Fifth column gives the nuisance parameter, $M$. Sixth column contains $\chi^2_{\textrm{min}}$ whilst seventh up to tenth column includes the statistical indicators and their difference with respect to $\Lambda$CDM model.}
    \label{tab:LM}
    \begin{tabular}{>{\centering}m{0.18\textwidth}>{\centering}m{0.1\textwidth}>{\centering}m{0.12\textwidth}>{\centering}m{0.09\textwidth}>{\centering}m{0.14\textwidth}>{\centering}m{0.08\textwidth}>{\centering}m{0.06\textwidth}>{\centering}m{0.06\textwidth}>{\centering}m{0.06\textwidth}>{\centering\arraybackslash}m{0.06\textwidth}}
        \hline
		\rule{0pt}{1.2 \baselineskip}\rule{0pt}{1.2 \baselineskip}Data Sets & $H_0$[\si{\km/ \s / \mega \parsec}] & $\Omega_{m,0}$ & $p_3 / 10^{-3}$ & $M$ & $\chi^2_\mathrm{min}$& AIC & BIC & $\Delta$AIC & $\Delta$BIC \\ [4pt]
		\hline
		\rule{0pt}{1.2 \baselineskip}CC + SN & $68.8\pm 1.9$ & $0.258^{+0.025}_{-0.024}$ & $ 5.9^{+10.4}_{-4.0}$ & $-19.370^{+0.059}_{-0.056}$ & 1042.08 & 1050.08 & 1070.02 & 2.35  & 7.33 \\[4pt] 
		$ \mathrm{CC + SN + R21}$ & $72.09^{+0.97}_{-0.98}$ & $0.243^{+0.021}_{-0.022}$ & $ 5.5^{+10.8}_{-3.4}$ & $-19.274\pm 0.028$ & 1046.93 & 1054.93 & 1074.87 & 2.48 & 7.46  \\ [4pt]
		$\mathrm{CC+ SN + F21}$ & $69.3\pm 1.3$ & $0.257\pm 0.022$ & $ 5.3^{+9.4}_{-3.2} $ & $-19.355^{+0.039}_{-0.036}$ & 1042.23 & 1050.23 & 1070.17 & 2.35 & 7.33\\ [4pt]
		\hline
		\rule{0pt}{1.2 \baselineskip}CC + SN + BAO & $66.0^{+1.3}_{-1.4}$ & $0.300\pm 0.014$ & $1.53^{+5.13}_{-1.00} $ & $-19.442^{+0.035}_{-0.049}$ & 1049.21 & 1057.21 & 1077.20 & 1.20 & 6.20 \\ [4pt]
		$\mathrm{CC + SN + BAO + R21}$ & $69.64^{+0.73}_{-0.76}$ & $0.310\pm 0.013$ & $ 0.32^{+0.19}_{-0.0} $ & $-19.340^{+0.023}_{-0.024}$ & 1066.77 & 1074.77 & 1094.77 & 5.33 & 10.33 \\ [4pt]
		$\mathrm{CC+ SN + BAO + F21}$ & $67.17\pm 0.94$ & $0.303^{+0.013}_{-0.012}$ & $ 0.49^{+1.7}_{-0.12}$ & $-19.429^{+0.047}_{-0.019}$ & 1049.78 & 1057.78 & 1077.78 & 1.09 & 6.09\\ [4pt]
		\hline
    \end{tabular}
\end{table}

%%%%%%%%%%%%%%%%%%%%%%%%%%%%%%%%%%%%%%%%%%%%%%%%
%%%%%%%%%%%%%%%%%%%%%%%%%%%%%%%%%%%%%%%%%%%%%%%%

\section{\label{sec:conc}Conclusion}

\begin{figure}[htbp]
    \centering
    \hbox{\hspace{-2 em}
    \includegraphics[width = \columnwidth]{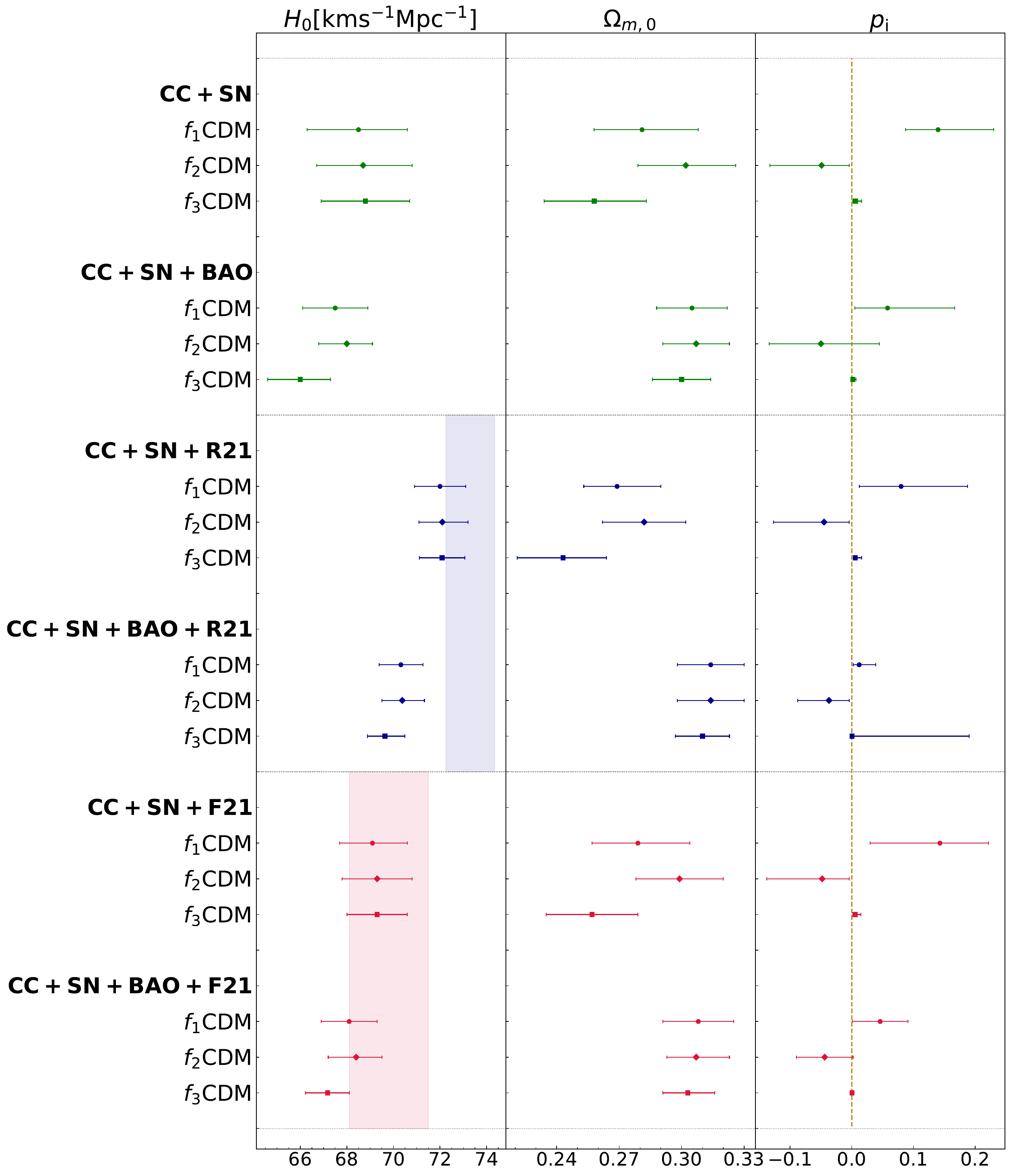}}
    \caption{Whisker plot for model parameters $H_0$, $\Omega_0$ and $p_i$ respectively. The additional parameter $p_i$ corresponds to: $p_1$ for $f_1$CDM, $p_2$ for $f_2$CDM and $p_3$ for $f_3$CDM. The best fits are reported for CC+SN and CC+SN+BAO as: without prior (green), with R21 prior (blue), and with F21 prior (red). The rectangles in colors denote the 1$\sigma$ uncertainties for each prior for the Hubble constant. The dashed orange line for the $p_i$ parameter represents the $\Lambda$CDM value for each mode (except the $f_3$CDM model).}
    \label{fig:whisker_plot}
\end{figure}

In our study, we probe the performance of three prominent models in $f(T,B)$ gravity against cosmological observations of expansion. We also survey how priors on the Hubble constant affect the posterior outcomes for the individual cases. This is done in the context of CC, SN and BAO data which are particularly relevant for the $H_0$ problem in the late Universe. Our main interest was to determine the impact that these combinations of analyses would have on the output cosmological parameters that each model produces, as compared with $\Lambda$CDM. In order to highlight the contrasts between these analyses, we show the whisker plot in Fig.~\ref{fig:whisker_plot}. Here, each prior on the Hubble constant is shown as a shaded region with its uncertainties, thus showing how each of the three models is individually impacted. Together with this, the other model parameters are shown, which do show some variance across the models. The appearance of the $H_0^{\rm F21}$ appears to be consistent with the models, in particular the CC+SN combination results in less than 1-sigma difference between the priors and posterior outputs, which rises slightly in the case that BAO data is included. On the other hand, the very high $H_0^{\rm R21}$ prior on the Hubble constant results in a mostly 2-sigma difference between these values, which is somewhat expected due to this being one of the highest literature values of $H_0$.

The first two models we explore in this work, $f_1$CDM and $f_2$CDM, host a continuous $\Lambda$CDM in which a specific setting of the additional model parameter recovers a constant cosmological constant contribution. In this way, we are able to assess whether the analysis prefers a nonzero evolution for the additional Lagrangian terms in Eqs.~(\ref{eq:PLM},\ref{eq:EM}). For comparison purposes, we also run the same analyses for the $\Lambda$CDM model putting the respective results in Appendix~\ref{sec:app}, which is useful for statistical comparisons. By and large, these models are consistent with the $\Lambda$CDM model. However, the $f_1$CDM model consistently allows for a lower value of the $\Omega_{m,0}$ parameter while the $f_2$CDM model increases this value, except for the case where BAO data is included where both models produce very similar values for the matter density parameter. On the other hand, the additional model parameters for these models at times goes marginally out of the 1$\sigma$ uncertainties but still being consistent with $\Lambda$CDM. This motivates further study using more data sets in cosmology or astrophysics. These results are also consistent with Ref.~\cite{Escamilla-Rivera:2019ulu} which also performed a background analysis on similar $f(T,B)$ models.

On the other hand, the $f_3$CDM model does not have a $\Lambda$CDM limit since no value of the $p_3$ parameter in Eq.~\eqref{eq:LogM} will make this term a constant. These models can be interesting since they do not alter $\Lambda$CDM in a marginal way but offer a genuine alternative to it. Interestingly, this model points to a lower value of Hubble parameter for most data set combinations, and similarly for the matter density parameter. Interestingly, the additional parameter for the $f_3$CDM model gives very small uncertainties for some of the data samples even at background level. This may indicate further analysis may yield decisive information about the model.

All three model appear to approximate the observational data reasonably well, as evidenced by the individual posterior plots in Figs.~\ref{fig:M1_CL}, \ref{fig:M2_CL}, \ref{fig:M3_CL}, as well as the whisker plot shown in Fig.~\ref{fig:whisker_plot}. Statistically, the $f_2$CDM model seems to mimic the priors best when they are present but this statement is well within the 1$\sigma$ uncertainties. Another property of all three models is that they agree with $\Lambda$CDM to with 1-2$\sigma$ for the most part, but they do give some freedom to better meet the observational demands of recent measurements. To further test this statement, we intend to probe the models further using early time data using the modified Boltzmann equations in a future study.

The work presented here explores the viability of literature models in $f(T,B)$ gravity together with their consistency with priors on the Hubble constant. It would be interesting to now extend this work to forecast data of future surveys, and also to consider CMB data from recent surveys such as the Planck Mission. This would however require the use of perturbations such as in Ref.~\cite{Bahamonde:2020lsm}. We intend to extend this work to study the early phases of the Universe in this scenario including an analysis on the effect that such models may have on inflationary scenarios, in a future work.

\appendix
\section{LCDM Model} \label{sec:app}

\begin{figure}[H]
    \centering
    \includegraphics[scale = 0.4]{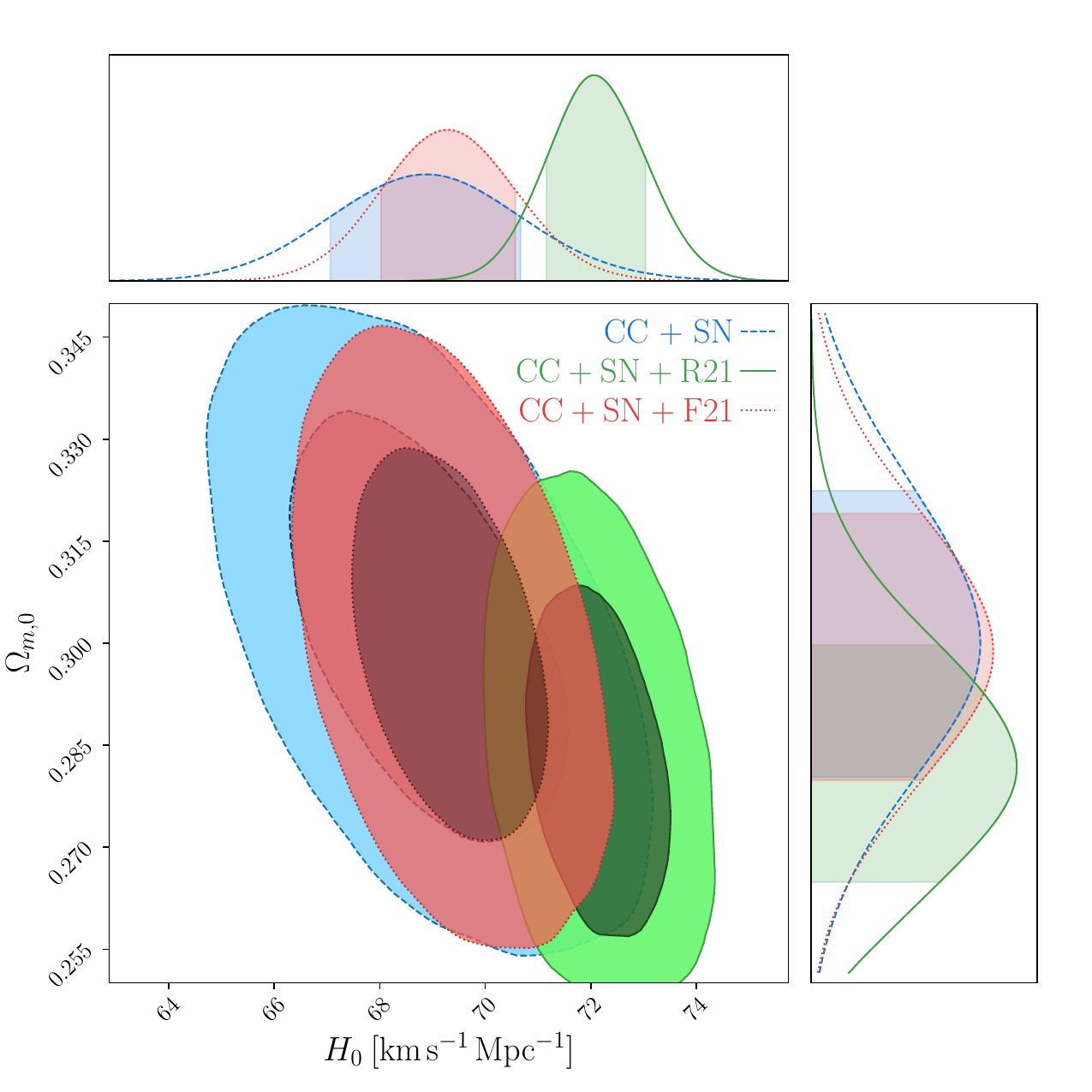}
     \includegraphics[scale = 0.4]{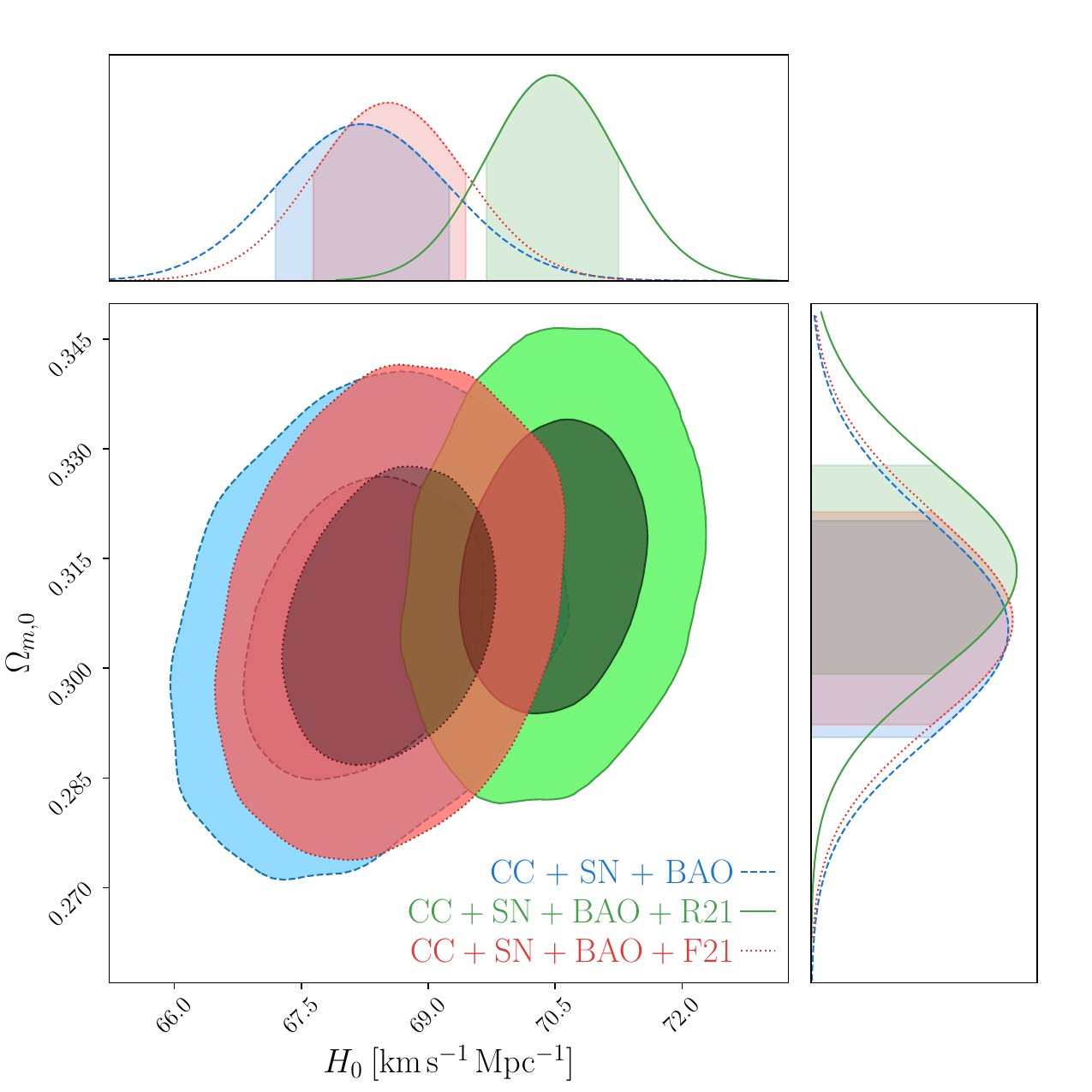}
    \caption{C.L for the $\Lambda$CDM model. The regions in color denotes: CC+SN sample (blue), CC+SN+R21 (green) and CC+SN+F21 (red). The priors R21 and F21 indicates the values quoted in Sec.~\ref{sec:obs_data}.}
    \label{fig:M0CCSN}
\end{figure}

In this appendix we present, for comparison purposes, the statistical results using the same analyses for the $\Lambda$CDM model (see Figure \ref{fig:M0CCSN}) and its statistical comparison between data samples in Table \ref{tab:LCDM}.

\begin{table}
\small
    \centering
    \caption{LCDM Model Caption}
    \label{tab:LCDM}
    \begin{tabular}{>{\centering}m{0.22\textwidth}X>{\centering}m{0.14\textwidth}>{\centering}m{0.14\textwidth}>{\centering}m{0.12\textwidth}>{\centering}m{0.1\textwidth}>{\centering\arraybackslash}m{0.1\textwidth}>{\centering\arraybackslash}m{0.1\textwidth}}
        \hline
		\rule{0pt}{1.2 \baselineskip}\rule{0pt}{1.2 \baselineskip}Data Sets & $H_0$[\si{\km/ \s / \mega \parsec}] & $\Omega_{m,0}$ & $M$ & $\chi^2_\mathrm{min}$ & AIC & BIC\\ [4pt]
		\hline
		\rule{0pt}{1.2 \baselineskip}CC + SN & $68.8\pm 1.8$ & $0.300^{+0.023}_{-0.019}$ & $-19.381^{+0.051}_{-0.054}$ & 1041.74 & 1047.74 & 1062.83 \\[4pt] 
		$ \mathrm{CC + SN + R21}$ & $72.05^{+0.99}_{-0.90}$ & $0.282^{+0.018}_{-0.017}$ & $-19.292^{+0.028}_{-0.027}$ & 1046.45 & 1052.45 & 1067.41 \\[4pt]
		$\mathrm{CC+ SN + F21}$ & $69.3\pm 1.3$ & $0.299^{+0.020}_{-0.019}$ & $-19.370\pm 0.038$ & 1041.88 & 1047.88 & 1062.83\\ [4pt]
		\hline
		\rule{0pt}{1.2 \baselineskip}CC + SN + BAO & $68.2\pm 1.0$ & $0.305\pm 0.015$ & $-19.403^{+0.037}_{-0.034}$ & 1050.01 & 1056.01 & 1071.00 \\ [4pt]
		$\mathrm{CC + SN + BAO + R21}$ & $70.46^{+0.80}_{-0.77}$ & $0.313\pm 0.014$ & $-19.328^{+0.027}_{-0.026}$ & 1063.44 & 1069.44 & 1084.45 \\ [4pt]
		$\mathrm{CC+ SN + BAO + F21}$ & $68.54\pm 0.90$ & $0.306^{+0.015}_{-0.014}$ & $-19.389^{+0.030}_{-0.032}$ & 1050.68 & 1056.68 & 1077.69\\ [4pt]
		\hline
    \end{tabular}
\end{table}

%%%%%%%%%%%%%%%%%%%%%%%%%%%%%%%%%%%%%%%%%%%%%%%%
%%%%%%%%%%%%%%%%%%%%%%%%%%%%%%%%%%%%%%%%%%%%%%%%
\begin{acknowledgments}
CE-R acknowledges the Royal Astronomical Society as FRAS 10147 and supported by PAPIIT Project TA100122. This work is part of the Cosmostatistics National Group (\href{https://www.nucleares.unam.mx/CosmoNag/index.html}{CosmoNag}) project. The authors would like to acknowledge networking support by the COST Action CA18108 and funding support from Cosmology@MALTA which is supported by the University of Malta. This research has been carried out using computational facilities procured through the European Regional Development Fund, Project No. ERDF-080 ``A supercomputing laboratory for the University of Malta''. The authors would also like to acknowledge funding from ``The Malta Council for Science and Technology'' in project IPAS-2020-007.
\end{acknowledgments}

%%%%%%%%%%%%%%%%%%%%%%%%%%%%%%%%%%%%%%%%%%%%%%%%
%%%%%%%%%%%%%%%%%%%%%%%%%%%%%%%%%%%%%%%%%%%%%%%%

\providecommand{\href}[2]{#2}\begingroup\raggedright\endgroup

\end{document}